\documentclass[aps,prd,twocolumn,superscriptaddress,nofootinbib]{revtex4-2}

\usepackage{tikz}
\tikzset{every picture/.style={line width=0.75pt}}
\usepackage{graphicx}
\usepackage{dcolumn}
\usepackage{bm}
\usepackage{tikz-feynman}

\usepackage{xcolor}
\definecolor{newtext}{RGB}{0,105,65}

\usetikzlibrary{decorations.markings,decorations.pathmorphing,shapes}
\usetikzlibrary{angles,quotes}
\usetikzlibrary{arrows.meta}
\tikzstyle{photon}=[line width=0.4,decorate, decoration={snake,amplitude=2,segment length=4,post length=0.1}]
\usepackage[italicdiff]{physics}
\usepackage[normalem]{ulem}

\usepackage[
    colorlinks=true, 
    citecolor=blue,   
    linkcolor=blue,   
    urlcolor=blue     
]{hyperref}

\begin{document}

\title{On the magnetic counterpart of the Uehling correction}


\author{T. Azevedo}
\email{thales@if.ufrj.br}
\affiliation{Instituto de F\'{\i}sica, Universidade Federal do Rio de Janeiro, Caixa Postal 68528, RJ 21941-972, Brazil}

\author{F.A. Barone}
\email{fbarone@unifei.edu.br}
\affiliation{IFQ - Universidade Federal de Itajub\'a, Av. BPS 1303, Pinheirinho, Caixa Postal 50, 37500-903, Itajub\'a, MG -- Brazil}

\author{C. Farina}
\email{farina@if.ufrj.br}
\affiliation{Instituto de F\'{\i}sica, Universidade Federal do Rio de Janeiro, Caixa Postal 68528, RJ 21941-972, Brazil}

\author{R. de Melo e Souza}
\email{reinaldos@id.uff.br}
\affiliation{Instituto de F\'{\i}sica, Universidade Federal Fluminense, Avenida Litor\^anea s/n, CP 24210-346 Niter\'oi, Rio de Janeiro -- Brazil}

\author{G. Zarpelon}
\email{gzarpelon0@gmail.com}
\affiliation{Instituto de F\'{\i}sica, Universidade Federal do Rio de Janeiro, Caixa Postal 68528, RJ 21941-972, Brazil}
\begin{abstract}
In this work, we investigate the magnetic properties of the quantum vacuum in the context of QED. We calculate the quantum relativistic correction of virtual particle-anti-particle pair creation to the field of a classical point-like idealized magnet. Using such correction, we find the induced currents stemming from the effect of the polarization of the vacuum, which behaves like a paramagnetic medium. We also calculate the correction to the electric dipole potential and show that the well-known symmetry between the classical fields of point electric and magnetic dipoles is broken at the quantum level. Lastly, to illustrate possible applications of our findings, we briefly discuss  the effect of the corrections obtained here in the context of neutron-neutron scattering and the hyperfine splitting for hydrogen-like atoms. 
\end{abstract}

\maketitle

\flushbottom


\section{\label{sec:level1}Introduction and Overview}

Quantum electrodynamics (QED) is a central tenet of modern physics and the prototype for any quantum field theory. A striking feature of this theory is the non-trivial properties of the quantum vacuum. The continuous creation and annihilation of virtual particle-anti-particle pairs translates into the behavior of the vacuum as a nonlinear material which can be polarized~\cite{dittrich,GreinerRafelski,GittmanBook,DittrichGies,MilonniVacuumBook,MiltonBook}. Recent experimental advances enabled direct observation of different nonlinear intriguing phenomena such as photon splitting~\cite{Akhmadaliev2002,di2007photon,Lotstedt2012}, photon-photon scattering~\cite{atlas1,Aad2019,Sirunyan2026} and vacuum birefringence~\cite{bragin2017high}. Those measurements in turn stimulate fundamental investigations on vacuum polarization, a topic driven by an active and intense community~\cite{Formanek2024,rafelskilandau2025,Volkov2025,Ramsey2026,Volkov2026,Dunne2026}. 

The first order correction due to the vacuum polarization to the Coulomb field produced by a point charge is known as the Uehling potential~\cite{uehling1935}. Since its original proposition, this potential has been extensively studied~\cite{wichmann1956,huang1976,Petelenz1987,frolov2012,Indelicato2014,greiner}, not only because of its intrinsic importance, but also since its knowledge is necessary in many situations, as for instance in the calculation of the Lamb shift~\cite{lamb1947,bethe1947,karshenboim1999polarization,yerokhin2019theory,Frugiuelea2022, Krachkov2023}. Nowadays, the Uehling correction plays a central role in the analysis of precise experimental tests for QED~\cite{Sanamyan2023,Maza2025} and in scattering phenomena~~\cite{Amundsen2021,Xu2022,Amundsen2024,Ghosh2025,Fontes2025} and has been applied for a deep understanding of the spectra of many-electron atoms~\cite{Kumar2022,Fairhall2023,Hasted2025}, ordinary molecules~\cite{Janke2025,Flynn2025}, and muonic and pionic molecules~\cite{Karr2013,Michel2019}. Recent theoretical developments have focused on establishing different representations and calculation  schemes~\cite{Burgess2017,barone2018,Frolov2021,Mohr2023,Frolov2024}, investigating higher-order corrections~\cite{Ivanov2024, Flynn2025b} and examining the influence of boundaries~\cite{nosso25}, to name a few. In addition, the analogues of Uehling's result to modified QED theories have been analyzed~ \cite{Ajamieh2024, Ajamieh2025}, as well as extensions to the hadronic case~\cite{breidenbach2022hadronic,Dizer2023} and to quantum gravity theories~\cite{Draper2020,Cordova2022,Jimu2025}.

In this paper we generalize Uehling's correction to magnetism, where the absence of monopoles indicates that the leading correction only appears in the next multipole, i.e. the dipole field. The interest in this analysis is based on the fact that the radiative corrections arising from QED also have important consequences for the magnetic properties of a given system. In its most fundamental form, we recall the famous correction to the gyromagnetic factor of the electron. The interaction of magnetic dipoles with external electromagnetic fields within the framework of quantum field theory is still an active field of research~\cite{Evans2020,Bruce2020,Rafelski2023}. Here we focus instead on the vacuum polarization corrections to the magnetic field produced by the point magnetic dipole itself. Our results suggest that the QED vacuum can be thought of as a medium with anisotropic paramagnetic behavior, thus shedding new light into the character of the vacuum.

 In addition, we generalize Uehling's original treatment to account for the vacuum polarization correction to the field produced by a point electric dipole. Searches for an intrinsic electric dipole moment of the electron are currently of great interest as sensitive probes of physics beyond the Standard Model~\cite{ACME2018, Roussy2023, Stuntz2024, Bause2025}. A comparison of the Uehling correction for the electric dipole with the one for a magnetic dipole reveals that the quantum vacuum breaks the symmetry present in Maxwell's theory between the fields produced by electric and magnetic dipoles---see Appendix~\ref{appendix:dipole_symmetry} for a first principles derivation of the classical symmetry. 

 Even though it is not the main goal of the present work, we briefly discuss two possible applications of our results. First,   we argue that the formalism presented here is directly relevant for future precision experiments in QED, particularly those probing the magnetic fields generated by subatomic particles with intrinsic magnetic dipole moments. Indeed, we show that the magnetic Uehling correction leaves a signature in the long-distance neutron-neutron interaction, a problem of interest in nuclear physics. Second, we employ the alternative approach developed here to revisit the Uehling magnetic correction to the hyperfine structure, which was treated in Ref.~\cite{Artemyev2001} and references therein.

The remainder of the paper is organized as follows. In section~\ref{section:first}, we set our notation/conventions and review the standard Uehling potential, as well as a straightforward extension to the electric dipole correction. In section~\ref{section:magnetic_case}, we present the calculation for the corrected magnetic dipole field. In section~\ref{section:para_vacuum}, we give an interpretation of the results in light of the induced vacuum charges and discuss the magnetic character of the vacuum.  In section~\ref{section:applications}, we  outline two possible applications of our findings, as explained above. 
Finally, we conclude with a few comments in section~\ref{secion:conclusion}. We adopt the mostly-negative metric signature, $\eta_{\alpha\beta} = \text{diag}(1,-1,-1,-1)$, with indices from the Greek alphabet running from $0$ to $3$, and express three-vectors in bold (with components labeled by Latin letters $j,k,\ldots$).


\section{The Quantum-Corrected Electric Dipole Field \label{section:first}}

In this section, we present the  calculation for the leading-order correction of the electric dipole field due to vacuum polarization. This section serves to set notation and also to establish a result to be compared further on. 

\subsection{Preliminaries and Setup}

We are concerned with the standard theory of Quantum Electrodynamics (QED), whose Lagrangian density is  
\begin{eqnarray}
    \mathcal{L}_{QED} = -\frac{1}{4}F_{\mu\nu} F^{\mu\nu} + \bar\psi(i\gamma^\mu D_\mu - m_f)\psi\,,
\end{eqnarray}
with $F_{\mu\nu} = \partial_{\mu}A_{\nu}-\partial_{\nu}A_{\mu}$, $A_\mu$ being the gauge (photon) field, $\psi$ being the fermion field of mass $m_f$ (with $\bar\psi := \gamma^0 \psi^\dagger$), and the covariant derivative $D_\mu := \partial_\mu - ieA_\mu$. In this context, we take the   photon propagator in momentum space to be
\begin{eqnarray}
    \Delta_{\mu\nu}(p) = \frac{-i}{p^2 + i\epsilon}\left(\eta_{\mu\nu} - \left(1-\xi\right)\frac{p_\mu p_\nu}{p^2} \right),
\end{eqnarray}
where $\epsilon$ is the constant associated with the Feynman prescription for propagators and $\xi$ is the gauge parameter. For the fermions of mass $m_f$ running inside the loop correction, we consider the propagators $G_F$ to be
\begin{eqnarray}
    G_F(p) = \frac{-i(\gamma^\mu p_\mu + m_f)}{p^2 - m^2 + i\epsilon}\,.
\end{eqnarray}

The vacuum polarization corrections to the photon propagator arise naturally in the perturbative expansion of QED, leading to the renormalization of the electric charge $e$. These corrections are encoded in the polarization tensor $\Pi^{\mu\nu}$, which is a transverse tensor that can be conveniently written as $\Pi^{\mu\nu} = (\eta^{\mu\nu}p^2 - p^\mu p^\nu) \Pi(p^2)$. The well-known renormalization process renders this (formally) infinite quantity finite, and yields the standard perturbative expansion in the fine structure constant $\alpha=e^2/(4\pi)$.
To translate the above statement into mathematical language, we note that the photon propagator can be written as
 \begin{eqnarray}
     \Delta^{(0)}_{\mu\nu} = \frac{-i}{p^2 + i\epsilon} P_{\mu\nu} + \frac{-i}{(p^2 + i\epsilon)^2}p_\mu p_\nu\xi\,,\nonumber 
 \end{eqnarray}
 where $P_{\mu\nu} = \eta_{\mu\nu} - p_\mu p_\nu/p^2$ corresponds to a projector, satisfying $P_{\mu\nu} P^{\nu\alpha} = P_\mu^{\ \ \alpha}$. Noting that $i\Pi^{\mu\nu} = ip^2 \Pi(p^2) P^{\mu\nu}$, we can write the $n$-th iteration of the bubble diagram as
 \begin{eqnarray}
    \Delta^{(0)}_{\mu\alpha_1} i\Pi^{\alpha_1 \beta_1} \Delta^{(0)}_{\beta_1\alpha_2}i\Pi^{\alpha_2 \beta_2} \dots i\Pi^{\alpha_n \beta_n} \Delta^{(0)}_{\beta_n\nu} \nonumber\\= \frac{1}{p^2 + i\epsilon} \left(-i\Pi(p^2)\right)^n P_{\mu\nu}\,. \nonumber\\
 \end{eqnarray}
 Then, the full 1PI diagrams are given by the series
 \begin{eqnarray}
     \Delta^{(1)}_{\mu\nu} &=& \Delta^{(0)}_{\mu\nu} +\frac{1}{p^2+i\epsilon} \sum_{n=1}^\infty \left(-i\Pi\right)^n P_{\mu\nu} \nonumber\\
     &=& \frac{-i}{p^2 + i\epsilon}P_{\mu\nu} \sum_{n=0}^\infty (-i)^{n-1}\Pi^n +\frac{-i}{(p^2+i\epsilon)^2}p_\mu p_\nu \xi \,.\nonumber\\
 \end{eqnarray}
 Thus, the multiple corrections can be incorporated by using the form above. The last piece is explicitly dependent on the gauge parameter $\xi$ and vanishes in Lorenz/Landau gauge with $\xi = 0$. 

The starting point of the calculation follows the prescription to extract the potential from  a classic source via the propagator of a given gauge field $A_\mu$. Namely,
\begin{eqnarray}\label{potencialfontes}
    iA^{(n)}_\mu(x) = \int \dd[4]{y} \Delta^{(n)}_{\mu\nu}(x-y) j^{\nu}(y)\,,
\end{eqnarray}
where the $(n)$ superscript indicates the loop order one is considering. For the leading contribution beyond tree level ($n=1$, often called the \textit{Uehling correction}~\cite{uehling1935,greiner,peskin}),  $\Delta^{(1)}_{\mu\nu}$ can be easily found. 


{The covariance of Quantum Electrodynamics allows us to work in the more convenient frame. Following Uehling~\cite{uehling1935}, we consider the case of a point charge whose rest frame is an inertial frame. Choosing the origin at the point charge position we have $j^{\nu}=q(\delta(\boldsymbol{r}),\boldsymbol{0})$. Substituting this 4-current into Eq.~(\ref{potencialfontes}) we obtain the well-known Uehling result}

\begin{eqnarray}
    \phi^{(1)}(r) &=& \frac{q}{4\pi r}\left[1 +\frac{2\alpha}{3\pi} \mathcal{I}_0(m_fr)\right]\!,\\ \mathcal{I}_n(z) &:=& \int_1^\infty\!\!\dd{\xi}U(\xi)  \xi^{-n}e^{-2\xi z}\,,
    \label{UehlingPotential1}
\end{eqnarray}
where we define 
\begin{eqnarray}
    U(\xi) := \left(1 + \frac{1}{2\xi^2}\right)\frac{\sqrt{\xi^2 -1}}{\xi^2}\,.
\end{eqnarray}

It is convenient to define an effective charge, to first order in $\alpha$, as
\begin{equation}
\label{defq1}
    q^{(1)}(r):= q \left[1 +\frac{2\alpha}{3\pi}    \mathcal{I}_0(m_fr)\right] \ ,
\end{equation}
so that the potential (\ref{UehlingPotential1}) takes the Coulomb-like form
\begin{equation}
\phi^{(1)}(r)=\frac{q^{(1)}(r)}{4\pi r}\ . 
\end{equation}


The integral $\mathcal{I}_0$ which appears in \eqref{UehlingPotential1} has been extensively studied in its multiple possible representations in terms of special functions. The application of   derivatives to $\mathcal{I}_n(z)$ follows a recurrence relation of the form
\begin{eqnarray}
     \mathcal{I}'_n(z) = -2\mathcal{I}_{n-1}(z)\,, \label{recurrence}
\end{eqnarray}
which will be useful later.

\subsection{Electric Dipole Correction}

As a straightforward application, one could consider what happens for a classical static electric dipole represented by the  current   $j^\mu(x) = (-\vb{d}\cdot\grad \delta^3(\vb{x}), \vb{0})$ \cite{BHBJP2010}. From Eq.~(\ref{potencialfontes}) we see that  the potential is  simply given by $\phi^{(1)}_{ed}=-\vb{d}\cdot\grad \phi^{(1)}(r)$, where $\phi^{(1)}$ is the Uehling potential \eqref{UehlingPotential1}. Explicitly,
\begin{eqnarray}
    \phi_{ed}^{(1)}(\vb{r}) &=&\frac{\vb{d}^{(1)}(r)\cdot\vb{r}}{4\pi r^3}\,,\label{phi_ed1}
\end{eqnarray}
where we have defined the renormalized dipole moment function $\vb{d}^{(1)}(r)$ as
\begin{eqnarray}
  &&\vb{d}^{(1)}(r):=
  \vb{d}\bigg[1 +\cr\cr
  &&+ \frac{2\alpha}{3\pi}\int_1^\infty \dd{\xi} U(\xi) ( 1 +2m_fr\xi)e^{-2m_fr\xi}\bigg]\,.
    \label{elec_dipolemom_renorm}
\end{eqnarray}
This can also be understood as the limiting case of the potential generated by two opposite point charges infinitesimally close to each other, corrected by vacuum polarization. Since $m_f\neq 0$ the renormalization factor for the electric dipole differs from that of the monopole case, given in Eq.~(\ref{UehlingPotential1}). This means that, although the superposition principle remains valid, there are non-trivial modifications produced by the vacuum polarization. In order to unveil them, let us evaluate the electric field produced by the electric dipole, given by $\vb{E}_{ed}^{(1)} = - \grad\phi_{ed}^{(1)}$. The correction to the field is given by
\begin{eqnarray}
    &&\delta\vb{E}_{ed}^{(1)}(\vb{r})= \frac{\alpha}{6\pi^2 r^3}\times\nonumber\\
    &&\times \Big[\big(3\vb{\hat{r}}(\vb{d}\cdot\vb{\hat{r}}) - \vb{d}\big)\big(\mathcal{I}_0(m_fr) + 2m_fr \mathcal{I}_{-1}(m_fr)\big) \nonumber\\
    &&+4m_f^2r^2\vb{\hat{r}}(\vb{d}\cdot\vb{\hat{r}})\mathcal{I}_{-2}(m_fr)\Big],
    \label{E_field_correction}
\end{eqnarray}
where we have used the recurrence relation \eqref{recurrence} with $z=m_fr$. The complete field becomes
\begin{eqnarray}
    \vb{E}_{ed}^{(1)}(\vb{r}) &=& \frac{1}{4\pi r^3}\Big(3\vb{\hat{r}}(\vb{d}^{(1)}(r)\cdot\vb{\hat{r}}) - \vb{d}^{(1)}(r)\Big)\nonumber\\
    && + \frac{\alpha}{6\pi^2 r^3} \vb{\hat{r}}(\vb{\hat{r}}\cdot\vb{d}) \int_1^\infty \dd{\xi} U(\xi) 4m_f^2 \xi^2 r^2 e^{-2m_fr\xi}\,. \nonumber\\\label{Electric_Dipole_Field} 
\end{eqnarray}

The first term was expected and  corresponds to the usual dipole electric field produced by the renormalized electric dipole. It is the only term surviving in the limit $\alpha\to 0$ (in which case $\mathbf{d}^{(1)}(r)\to \mathbf{d}$). The second term constitutes a non-trivial contribution arising due to the vacuum polarization, with a tensorial structure differing from the dipole electric field in classical electrodynamics. 

There is another interesting consequence of breaking the scale symmetry. For a given radial direction, the electric field direction changes as a function of the distance, that is, for $m_f\neq 0$ we have
\begin{equation}
\frac{d}{dr}\frac{\mathbf{E}_{ed}^{(1)}\cdot\boldsymbol{\hat{r}}}{|\mathbf{E}_{ed}^{(1)}|} \neq 0 \, .    
\end{equation}
In classical electromagnetism, however, dimensional analysis by itself ensures that this derivative vanishes for all point multipoles.

We stress that Eq.~(\ref{Electric_Dipole_Field}) is only valid   for $\mathbf{r}\neq \mathbf{0}$. To include the origin, one could apply a regularization procedure~\cite{farinamexicano} to the potential in Eq.~(\ref{phi_ed1}) as follows. Taking $\phi^{(1)}_{ed} \to \phi^{(1)}_{ed(\epsilon)}$, we write
\begin{eqnarray}
   \phi^{(1)}_{ed(\epsilon)} = -\frac{1}{4\pi}\vb{d}^{(1)}(r)\cdot \grad\left(\frac{1}{\sqrt{r^2 + \epsilon^2}}\right)\,.
\end{eqnarray}
Then, after calculating the gradient of the above, the limit $\epsilon\to 0$ is to be taken with the understanding that the term proportional to $\epsilon^2/(r^2 + \epsilon^2)^{5/2}$ yields a  Dirac delta function. This leads to 
\begin{eqnarray}
&&   \vb{E}^{(1)}_{ed}(\vb{r}) = \frac{1}{4\pi r^3} \Big(3\vb{\hat{r}}(\vb{d}^{(1)}(r)\cdot\vb{\hat{r}}) - \vb{d}^{(1)}(r)\Big)  +\nonumber\\
   &-& \frac{1}{3}\vb{d}^{(1)}(0) \delta(\vb{r}) + \frac{\alpha}{6\pi^2 r^3} 4m_f^2 \vb{r}(\vb{r}\cdot \vb{d}) \mathcal{I}_{-2}( m_fr)\,.
   \label{electric_dipole_withdelta}
\end{eqnarray}

Note that the term proportional to the   delta function is the same as the one in classical electrodynamics but with $\mathbf{d}$ replaced by $\mathbf{d}^{(1)}$, as expected.

\section{The Magnetic Case\label{section:magnetic_case}}

Having discussed the leading $\alpha$ correction to the electrostatic fields of a point charge and a point electric dipole, our goal in this section is to investigate the magnetic counterpart of Uehling's results. Given that there is no magnetic monopole in Maxwell's theory, we will be interested in the leading quantum correction to the magnetic field of a point magnetic dipole, i.e. the lowest-order multipole available.

Furthermore, as is well-known, the magnetic dipole field has a very similar form to the electric dipole field in Maxwell's theory---see Appendix~\ref{appendix:dipole_symmetry} for a discussion on why this is the case. Hence, a natural question arising in this context is whether or not this similarity remains true for the first perturbative corrections which comes from vacuum polarization. Our calculations will allow us to settle this question.

For a static magnetic dipole at position $\vb{x}_s$, the current density is given by $j^\mu = (0,\vb{J})$, where $\vb{J}(\vb{x}) = -\vb{m}\times\grad \delta^3(\vb{x}-\vb{x}_s)$. By inserting this into the potential, one has
\begin{eqnarray}
    iA^{(1)}_{md,\mu}(x) = \int\dd[4]{y} \Delta^{(1)}_{\mu j}(x - y) [-\vb{m}\times\grad\delta^3(\vb{y}-\vb{x}_s)]^j  \nonumber\,.
\end{eqnarray}
Using the momentum representation of the propagator, one can  exchange the order of integrations to obtain
\begin{eqnarray}
    iA^{(1)}_{md,\mu}(x) &=& \int\frac{\dd[4]{p}}{(2\pi)^4} \Delta^{(1)}_{\mu j}(p) e^{-ip\cdot x}\nonumber\\
    &&\times\int \dd[4]y [-\vb{m}\times\grad\delta^3(\vb{y}-\vb{x}_s)]^j e^{ip\cdot y}\,.\nonumber
\end{eqnarray}
However, since $\grad \delta^3(\vb{y} - \vb{x}_s) = -\grad_{\vb{x}_s} \delta^3(\vb{y} - \vb{x}_s)$, we can take the cross product out   of the integrals and write 
\begin{eqnarray}
    iA^{(1)}_{md,\mu}(x)
    = (\vb{m}\times\grad_{\vb{x}_s})^j \int\frac{\dd[4]{p}}{(2\pi)^3} \Delta^{(1)}_{\mu j}(p) e^{-ip\cdot x-i\vb{p}\cdot\vb{y}} \delta(p^0)\,. \nonumber
\end{eqnarray}
When the delta is integrated, the $p^0 $ components are all set to zero. This implies that $\Delta^{(1)}_{0j} = 0$, and we immediately see that $A^{(1)}_{md,0} = 0$, as expected. Thus, we have
\begin{eqnarray}
    iA^{(1)}_{md,k} (x) =  (\vb{m}\times\grad_{\vb{x}_s})^j\int\frac{\dd[3]{\vb{p}}}{(2\pi)^3} \Delta^{(1)}_{jk}(0, \vb{p})  e^{i\vb{p}\cdot(\vb{x} - \vb{x}_s)}\,.
\end{eqnarray}
Now, using the explicit form of the propagator, we can write
\begin{eqnarray}
     \Delta^{(1)}_{jk}(0, \vb{p}) &=& \Delta^{(0)}_{jk}(0,\vb{p}) + \delta\Delta^{(1)}_{jk}(0,\vb{p})\,,\nonumber\\
     &=& \frac{i}{\vb{p}^2 -i\epsilon}\left[\eta_{jk} - (1-\xi)\frac{p_j p_k}{-\vb{p}^2 + i\epsilon} \right]\nonumber\\
     &&+\frac{i}{\vb{p}^2 - i\epsilon} \Pi_R(-\vb{p}^2)\left[\eta_{jk} - \frac{p_j p_k}{-\vb{p}^2 + i\epsilon} \right]\nonumber\\
\end{eqnarray}
and it is clear that the components with $\mu =0$ of $\Delta^{(1)}_{\mu k}$ vanish, which is why we have swapped $\Delta^{(1)}_{\mu j} \to \Delta^{(1)}_{jk}$.

To organize the calculations, we can separate the two contributions as:
\begin{eqnarray}
    A^{(1)}_{md,j}(\vb{x}) = A^{(0)}_{md,j} (\vb{x}) + \delta A^{(1)}_{md,j}(\vb{x})\,,
\end{eqnarray}
with
\begin{align}
    iA^{(0)}_{md,j}(\vb{x}) =&\, (\vb{m}\times\grad_{\vb{x}_s})^k\int\frac{\dd[3]{\vb{p}}}{(2\pi)^3} \Delta^{(0)}_{jk}(0,\vb{p})  e^{i\vb{p}\cdot(\vb{x} - \vb{x}_s)}\,,\\
    i\delta A^{(1)}_{md,j}(\vb{x}) =& \, (\vb{m}\times\grad_{\vb{x}_s})^k \int\frac{\dd[3]{\vb{p}}}{(2\pi)^3} \delta\Delta^{(1)}_{jk}(0,\vb{p}) e^{i\vb{p}\cdot(\vb{x} - \vb{x}_s)}\,.
\end{align}
Note that this specific form of separating the contributions leaves the gauge parameter completely inside  the \textit{tree level} contribution, which is explicitly given by
\begin{align}
    A^{(0)}_{md,j}(\vb{x}) =  (\vb{m}\times\grad)_k &\int\frac{\dd[3]{\vb{p}}}{(2\pi)^3} \frac{e^{i\vb{p}\cdot(\vb{x} - \vb{x}_s)}}{-\vb{p}^2 + i\epsilon}\nonumber\\
    &\times\left[\delta_{jk} - (1-\xi) \frac{p_j p_k}{\vb{p}^2} \right] \,.
\end{align}

A few comments are in order. First, we have exchanged the derivative with respect to $\vb{x}_s$ by derivatives with respect to $\vb{x}$, which is justified because the dependence on space parameters is of the form $\vb{x}-\vb{x}_s$. Furthermore, since now there are no temporal components in the variables, we will abdicate the use of Lorentz-covariant notation and thus circumvent the appearance of extra minus signs which would plague us. By this we mean  that repeated indices  (whether up or down) are summed and $p^k = p_k$, formally.

We evaluate now the zeroth-order dipole field. The evaluation is straightforward. Let us focus on the integral. The first part is immediate,
\begin{eqnarray}
    \delta_{jk}\int\frac{\dd[3]{\vb{p}}}{(2\pi)^3} \frac{e^{i\vb{p}\cdot(\vb{x}-\vb{x}_s)}}{-\vb{p}^2 + i\epsilon} \overset{\epsilon\to0}{=} -\frac{\delta_{jk}}{4\pi r}\,,  
\end{eqnarray}
where $\vb{r}:= \vb{x} - \vb{x}_s$. The second piece can be written as 
\begin{eqnarray}
    (1-\xi)\partial_j\partial_k\int\frac{\dd[3]{\vb{p}}}{(2\pi)^3} \frac{e^{i\vb{p}\cdot(\vb{x}-\vb{x}_s)}}{\vb{p}^2(-\vb{p}^2 + i\epsilon)}\nonumber\\=
    \frac{(1-\xi)}{2\pi^2}\partial_j\partial_k\left[\frac{1}{r}\int_0^\infty\frac{p\sin(p r)}{p^2(-p^2 + i\epsilon)}\dd{p}\right].
\end{eqnarray}
The integral is straightforward by using complex integration. After taking $\epsilon \to 0$, one finds 
\begin{eqnarray}
    \frac{(1-\xi)}{8\pi}\partial_j\partial_kr = \frac{(1-\xi)}{8\pi r}\left(\delta_{jk} - \frac{r_j r_k}{r^2}\right).
\end{eqnarray}
Thus, the complete tree-level piece is
\begin{eqnarray}
    A^{(0)}_{md, j} = -(\vb{m}\times\grad)_k\left\{\frac{\delta_{jk}}{4\pi r}  - \frac{(1-\xi)}{8\pi r}\left( \delta_{jk} - \frac{r_j r_k}{r^2}\right)\right\}.
\end{eqnarray}
Operating the derivatives, one arrives at
\begin{eqnarray}
     A^{(0)}_{md,j}  &=& \frac{(\vb{m}\times\vb{r})_j}{4\pi r^3}\,.
\end{eqnarray}
We see that the gauge-dependent piece vanishes as expected for static sources. The field is then simply
\begin{eqnarray}\label{b0}
    B^{(0)i}_{md} = \frac{1}{4\pi r^3}\left(\frac{3r^i(\vb{r}\cdot\vb{m})}{r^2} - m^i\right),
\end{eqnarray}
which is the well-known classical result for an idealized point-like magnet.

The loop contribution is given by
\begin{align}
    i\delta A^{(1)}_{md,i}(\vb{x}) = i(\vb{m}\times\grad)_j&\int\frac{\dd[3]{\vb{p}}}{(2\pi)^3} \frac{e^{i\vb{p}\cdot(\vb{x} - \vb{x}_s)}}{-\vb{p}^2 + i\epsilon}\nonumber\\&\times \Pi_R(-\vb{p}^2)\left(\delta_{ij} - \frac{p_i p_j}{\vb{p}^2}\right) \,.
\end{align}
Thus, because of the projector inside the integral, we do not recover $(\vb{m}\times\grad)^i$ applied to the point particle Uehling correction. To pursue the above integration, we can split the integral in two contributions and deal with each piece of the projector separately. The first one, which is proportional to $\delta_{ij}$, yields the standard Uehling expression. To integrate the second term, we take $\epsilon\to 0$ and write
\begin{align}
    \int\frac{\dd[3]{\vb{p}}}{(2\pi)^3} \frac{\Pi_R(-\vb{p}^2)}{-\vb{p}^4} p_i p_j e^{i\vb{p}\cdot \Delta\vb{x}}\nonumber\\ = -\partial_i \partial_j\int\frac{\dd[3]{\vb{p}}}{(2\pi)^3} \frac{\Pi_R(-\vb{p}^2)}{-\vb{p}^4} e^{i\vb{p}\cdot \Delta\vb{x}}\,.\label{eq:3.16}
\end{align}
Using the form of the polarization tensor, it is straightforward to perform the above integration (see Appendix~\ref{apendix1}). The full integral is then given by 

\begin{eqnarray}
    &&\int\frac{\dd[3]{\vb{p}}}{(2\pi)^3} \frac{\Pi_R(-\vb{p}^2)}{-\vb{p}^2} \left(\delta_{ij} - \frac{p_i p_j}{\vb{p}^2}\right) e^{i\vb{p}\cdot(\vb{x} - \vb{x}_s)} \nonumber\\
    &=& -\frac{\alpha}{6\pi^2 r}\left(\delta_{ij} - \frac{r_i r_j}{r^2}\right) \int_1^\infty \dd{\xi} U(\xi) e^{-2m_f\xi r} \nonumber\\
    && -\frac{\alpha}{24m_f^2\pi^2 r^3}\left(\frac{3r_i r_j}{r^2} - \delta_{ij}\right)\nonumber\\
    &&\times\left[\frac{2}{5} - \int_1^{\infty} \dd{\xi} U(\xi)\left(\frac{1 + 2m_f\xi r}{\xi^2}\right) e^{-2m_f\xi r}\right]. \label{eq:3.17}
\end{eqnarray} 
At last, the correction $\delta A^{(1)}_i(\vb{x})$ is be given by applying $(\vb{m}\times\grad)_j$ to the expression above. Using the recurrence relations for the integrals, one can write the correction as
\begin{widetext}
\begin{eqnarray}
    \delta A^{(1)}_{md,j} &=& \frac{\alpha}{6\pi^2}\varepsilon_{jkl} m_k\left\{\frac{r_l}{r^3}\left(\delta_{ij} - \frac{r_i r_j}{r^2} \right)\mathcal{I}_0(m_fr) + \frac{1}{r^3}\left( \delta_{il}
    r_j + r_i \delta_{jl} -\frac{2r_i r_j r_l}{r^2}\right)\mathcal{I}_0(m_fr)\right.\nonumber\\
    && +\left(\delta_{ij} - \frac{r_i r_j}{r^2}\right)\frac{2m_fr_l}{r^2} \mathcal{I}_{-1}(m_fr) + \frac{3r_l}{4m_f^2 r^5}\left(\delta_{ij} - \frac{3r_i r_j}{r^2}\right)\left[-\frac{2}{5} + \mathcal{I}_{-2}(m_fr) + 2m_fr \mathcal{I}_{-1}(m_fr) \right]\nonumber\\
    &&+  \frac{3}{4m_f^2r^5}\left(-\frac{2r_ir_jr_l}{r^2} + \delta_{il}r_j + r_i \delta_{jl}\right)\left[-\frac{2}{5} + \mathcal{I}_{-2}(m_fr) + 2m_fr \mathcal{I}_{-1}(m_fr) \right]\nonumber\\
    &&\left.-\frac{r_l}{r^3}\left(\frac{3r_i r_j}{r^2} - \delta_{ij}\right)\mathcal{I}_0(m_fr)\right\}.
\end{eqnarray}
\end{widetext}
Some simplifications of this expression can be carried out by noting the contractions with the overall Levi-Civita symbol. There are effectively four types of contractions in the above expression. Two of them vanish by the anti-symmetry of $\varepsilon_{ijk}$. The surviving terms finally yield
\begin{eqnarray}
    \delta A^{(1)}_{md,j}(\vb{x}) =\frac{\alpha(\vb{m}\times\vb{r})_j}{6\pi^2r^3}\left(\mathcal{I}_0(m_fr) + 2m_fr \mathcal{I}_{-1}(m_fr) \right)\,.
\end{eqnarray}
Equivalently, in three-vector notation, we find that the vector potential is
\begin{eqnarray}\label{deltaAmd1}
    \delta\vb{A}_{md}^{(1)}(\vb{r}) = \alpha \frac{(\vb{m}\times\vb{r})}{6\pi^2r^3} \int_1^{ \infty} \dd{\xi} U(\xi)(1 + 2m_f\xi r) e^{-2m_f\xi r}\,.
\end{eqnarray}
%
%
Thus, the full vector potential, including the classical dipole piece becomes
\begin{align}
    \vb{A}_{md}^{(1)}(\vb{r}) &= \frac{\vb{m}\times\vb{r}}{4\pi r^3}\Big(1+\nonumber\\
    & +\frac{2\alpha}{3\pi}\int_1^\infty \dd{\xi} U(\xi) (1 + 2m_f\xi r)e^{-2m_f\xi r}\Big) \,. \label{dipole_potential1}
\end{align}

It is interesting to note that, up to this point, the structure of the correction follows closely the electric case.
Now, to find the effective fields corresponding to this potential, we recall the usual formulas $\vb{E}_{md} = -\partial_t \vb{A}_{md}$ and $\vb{B}_{md} = \grad\times\vb{A}_{md}$. Clearly, $\vb{E}_{md} = \vb{0}$, while $\vb{B}_{md}$ will be given by the curl of the above expression. We have\footnote{For completeness, we give representations of the  corrections  $\delta \vb{B}_{md}^{(1)}$ and $\delta \vb{E}_{ed}^{(1)}$ in terms of integrals of modified Bessel functions in Appendix~\ref{analytic}.}
\begin{eqnarray}
    \delta\vb{B}_{md}^{(1)}(\vb{r}) &=& \frac{\alpha}{6\pi^2r^3}\Big\{\Big(\mathcal{I}_0(m_fr) + 2m_fr \mathcal{I}_{-1}(m_fr)\Big)\nonumber\\
    &&\times\Big(3\vb{\hat{r}}(\vb{m}\cdot\vb{\hat{r}}) -\vb{m} \Big)\nonumber\\
    &&+ 4m_f^2 r^2\mathcal{I}_{-2}(m_fr)\Big(\vb{\hat{r}}(\vb{m}\cdot\vb{\hat{r}}) - \vb{m}\Big)\Big\}. \label{B_field_correction}
\end{eqnarray}
Including this piece, the full magnetic field becomes
\begin{align}
    &&\vb{B}_{md}^{(1)}(\vb{r}) = \frac{1}{4\pi r^3}\left(3\vb{\hat{r}} (\vb{m}^{(1)}(r)\cdot\vb{\hat{r}}) -\vb{m}^{(1)}(r)\right)\nonumber\\
    && +\frac{\alpha}{6\pi^2 r^3}\vb{\hat{r}}\times(\vb{\hat{r}}\times\vb{m}) \int_1^\infty\dd{\xi}U(\xi) 4m_f^2\xi^2r^2 e^{-2m_f r\xi}\,, \label{Magnetic_dipole_field}
\end{align}
where we defined the renormalized dipole moment $\vb{m}^{(1)}(r)$ as
\begin{eqnarray}
    &&\vb{m}^{(1)}(r) := \vb{m}\bigg[1 +\cr\cr
    &&+\frac{2\alpha}{3\pi}\int_1^\infty \dd{\xi} U(\xi)(1 + 2m_fr\xi) e^{-2m_fr\xi}\bigg]\,.
    \label{renormalized_magnetic_moment}
\end{eqnarray}

In order to find an expression valid everywhere, including $\vb{r}=\vb{0}$, one can carry out the same regularization procedure as done for the electric dipole, namely writing the vector potential as
\begin{eqnarray}
    \vb{A}^{(1)}_{md[\epsilon]}(\vb{r}) = -\frac{1}{4\pi} \vb{m}^{(1)}(r) \times \grad\left(\frac{1}{\sqrt{r^2 + \epsilon^2}}\right) 
\end{eqnarray}
and evaluating the magnetic field. The result, upon inclusion of the  delta function in the limit $\epsilon \to 0$, is 
\begin{eqnarray}
    &&\vb{B}_{md}^{(1)}(\vb{r}) = + \frac{2}{3} \vb{m}^{(1)} (0)\delta(\vb{r})\nonumber\\
    &&+\frac{1}{4\pi r^3}\left(3\vb{\hat{r}} (\vb{m}^{(1)}(r)\cdot\vb{\hat{r}}) -\vb{m}^{(1)}(r)\right)  \nonumber\\
    && +\frac{\alpha}{6\pi^2 r^3}\vb{\hat{r}}\times(\vb{\hat{r}}\times\vb{m}) \int_1^\infty\dd{\xi}U(\xi) 4m_f^2\xi^2r^2 e^{-2m_f r\xi}\,.\cr\cr
    &&\ \label{Magnetic_dipole_field_withdelta}
\end{eqnarray}
%
{We outline possible applications of the above expressions in section~\ref{section:applications}.}

As we see, Eqs.~\eqref{Electric_Dipole_Field} and \eqref{Magnetic_dipole_field} do not follow the classical symmetry which the dipole fields posses under the exchange of $\vb{d} \leftrightarrow \vb{m}$. Furthermore, as expected, when the limit $m_f \to \infty$ is taken, both fields reduce to the classical Maxwell theory and the symmetry for $r\neq 0$ is restored. 

\begin{figure*}[t]
    \centering
    \includegraphics[width=0.9\linewidth]{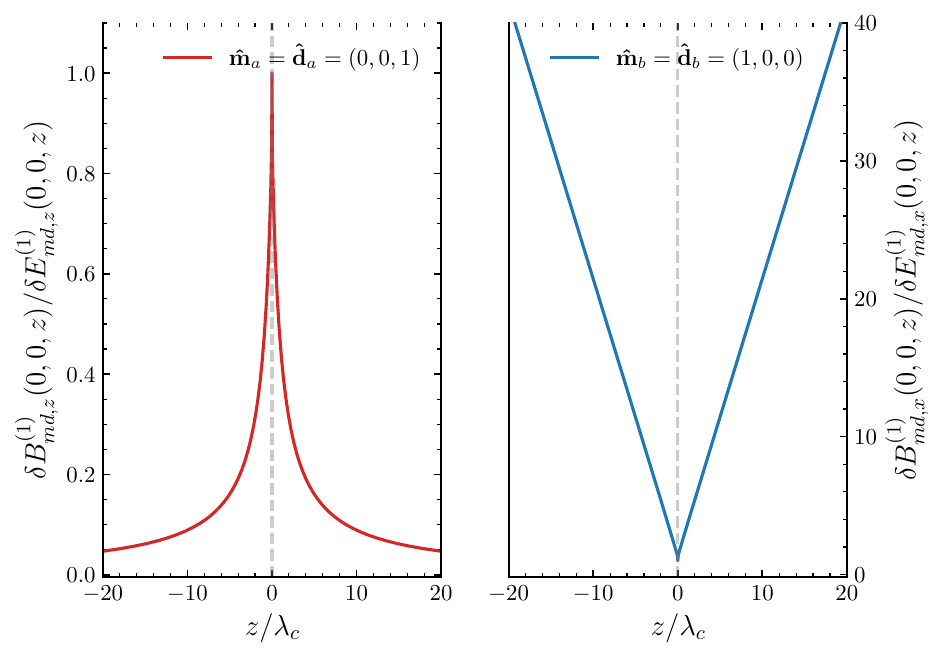}
    \caption{In the left panel, we plot $\frac{\delta \mathbf{B}_{md}(0,0,z)\cdot\mathbf{\hat{z}}}{\delta \mathbf{E}_{md}(0,0,z)\cdot\mathbf{\hat{z}}}$ with both dipoles in the $z$ direction. In the right panel, we plot $\frac{\delta \mathbf{B}_{md}(0,0,z)\cdot\mathbf{\hat{x}}}{\delta \mathbf{E}_{md}(0,0,z)\cdot\mathbf{\hat{x}}}$ with both dipoles in the $x$ direction. In both cases, the ratios are plotted as functions of $z$ normalized by the fermion Compton wavelength $(\lambda_c$). }       
    \label{fig:dipole_corr_ratios}
\end{figure*}

For comparison, and in order to highlight the breaking of the correspondence $\vb{d} \leftrightarrow \vb{m}$ in the quantum corrections to the dipole fields, we show in figure~\ref{fig:dipole_corr_ratios} plots of the ratios ($\delta B_{md,z}^{(1)}(0,0,z)/\delta E_{ed,z}^{(1)}(0,0,z)$ and $\delta B_{md,x}^{(1)}(0,0,z)/\delta E_{ed,x}^{(1)}(0,0,z)$), measured along the $z$ axis, for different dipole moment orientations. 

\section{Vacuum Magnetization \label{section:para_vacuum}}

In this section we shall investigate the magnetic character of the vacuum. This can be done by employing Maxwell's equations for the fields and identifying the sources which produce them. Due to the polarization of the vacuum, the electric dipole induces a polarization field while the magnetic dipole induces a magnetization field. The purpose of the section is to calculate these fields. 
If one computes the divergence of $\vb{B}_{dm}^{(1)}$, one finds $\grad\cdot \vb{B}_{dm}^{(1)} = 0$, as expected. On the other hand, the divergence of $\delta\vb{E}^{(1)}$ yields
\begin{eqnarray}
    \grad\cdot\delta\vb{E}_{de}^{(1)}(\vb{r}) &=& - \frac{\alpha}{6\pi^2}\frac{\vb{d}\cdot\vb{\hat{r}}}{r^4}
    \big[(2m_fr)^2 \mathcal{I}_{-2}(m_fr)\nonumber\\
    &&+ (2m_fr)^3 \mathcal{I}_{-3}(m_fr)\big].
\end{eqnarray}
For $r\neq 0$ the right-hand-side can be interpreted as the induced charge density, $\rho_b$, by a point dipole placed in the vacuum, up to order $\alpha$. By writing it as $\rho_b=-\boldsymbol{\nabla}\cdot\boldsymbol{P}$ we can identify the polarization field.
Furthermore, as a consistency check, $\grad\times\delta\vb{E}_{de}^{(1)} = \vb{0}$ (which is necessary for static sources). 

\begin{figure}[!h]
    \centering
    
\begin{tikzpicture}[x=0.75pt,y=0.75pt,yscale=-1,xscale=1]

\draw   (106,190.14) .. controls (106,154.79) and (134.65,126.14) .. (170,126.14) .. controls (205.35,126.14) and (234,154.79) .. (234,190.14) .. controls (234,225.48) and (205.35,254.14) .. (170,254.14) .. controls (134.65,254.14) and (106,225.48) .. (106,190.14) -- cycle ;
\draw   (78,190.14) .. controls (78,139.33) and (119.19,98.14) .. (170,98.14) .. controls (220.81,98.14) and (262,139.33) .. (262,190.14) .. controls (262,240.95) and (220.81,282.14) .. (170,282.14) .. controls (119.19,282.14) and (78,240.95) .. (78,190.14) -- cycle ;
\draw   (37,190.14) .. controls (37,116.68) and (96.55,57.14) .. (170,57.14) .. controls (243.45,57.14) and (303,116.68) .. (303,190.14) .. controls (303,263.59) and (243.45,323.14) .. (170,323.14) .. controls (96.55,323.14) and (37,263.59) .. (37,190.14) -- cycle ;
\draw    (303,190.14) -- (301.49,209.14) ;
\draw [shift={(301.33,211.14)}, rotate = 274.54] [color={rgb, 255:red, 0; green, 0; blue, 0 }  ][line width=0.75]    (10.93,-3.29) .. controls (6.95,-1.4) and (3.31,-0.3) .. (0,0) .. controls (3.31,0.3) and (6.95,1.4) .. (10.93,3.29)   ;
\draw    (262,190.14) -- (260.52,206.14) ;
\draw [shift={(260.33,208.14)}, rotate = 275.29] [color={rgb, 255:red, 0; green, 0; blue, 0 }  ][line width=0.75]    (10.93,-3.29) .. controls (6.95,-1.4) and (3.31,-0.3) .. (0,0) .. controls (3.31,0.3) and (6.95,1.4) .. (10.93,3.29)   ;
\draw    (234,190.14) -- (232.54,204.15) ;
\draw [shift={(232.33,206.14)}, rotate = 275.95] [color={rgb, 255:red, 0; green, 0; blue, 0 }  ][line width=0.75]    (10.93,-3.29) .. controls (6.95,-1.4) and (3.31,-0.3) .. (0,0) .. controls (3.31,0.3) and (6.95,1.4) .. (10.93,3.29)   ;

\draw (158,179.54) node [anchor=north west][inner sep=0.75pt]    {$\bigodot \ \mathbf{m}$};
\draw (304.17,204.04) node [anchor=north west][inner sep=0.75pt]    {$\ \boldsymbol{\mathcal{J}}^{(1)}$};

\end{tikzpicture}

 \caption{Lines of current density correction $\boldsymbol{\mathcal{J}}^{(1)}$ in the system's plane of symmetry.}
    \label{Fig:j}
\end{figure}
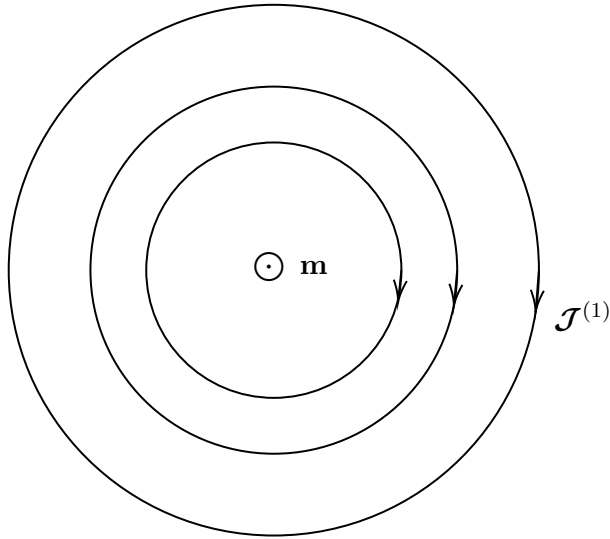

Concerning the magnetic field, calculating  the curl of its  correction yields
\begin{eqnarray}
    \grad\times\delta\vb{B}_{dm}^{(1)} &=& \frac{\alpha}{6\pi^2}\frac{\hat{\vb{r}}\times\vb{m}}{r^4}
    \big[(2m_fr)^2\mathcal{I}_{-2}(m_fr) \nonumber\\
    &&+ (2m_fr)^3 \mathcal{I}_{-3}(m_fr)\big].
    \label{eq:j_corrected}
\end{eqnarray}
Note that, if we interpret the right-hand side of Eq.~(\ref{eq:j_corrected}) as an $\alpha$-correction to the current density induced by the magnetic point dipole, $\boldsymbol{\mathcal{J}}^{(1)}$, then a field line representation of that current in the plane perpendicular to the magnetic dipole is shown in Figure~\ref{Fig:j}.

In order to analyze the magnetic character of the vacuum it is more convenient to deal with the magnetization field, defined by $\boldsymbol{\mathcal{J}}^{(1)}=\boldsymbol{\nabla}\times\boldsymbol{\mathcal{M}}^{(1)}$.  A natural choice for the magnetization is precisely $\boldsymbol{\mathcal{M}}^{(1)}=\delta\mathbf{B}^{(1)}$, given in Eq.~(\ref{B_field_correction}). The role of the first term  is to partially shield the magnetic field produced by the source $\mathbf{m}^{(1)}(0)$ placed at the origin.  The other term is a nontrivial contribution arising from vacuum polarization.

The magnetization can be written as 
\begin{equation}
    \mathcal{M}_i^{(1)} = (\Lambda_{ik}^{(a)}+\Lambda_{ik}^{(b)})m_k \, ,
\end{equation}
where we have defined the second rank tensors
\begin{align}
    \Lambda_{ik}^{(a)}&=\,\frac{\alpha\Big(\mathcal{I}_0(m_fr) + 2m_fr \mathcal{I}_{-1}(m_fr)\Big)}{6\pi^2r^3}(3\hat{r}_i\hat{r}_k-\delta_{ik}) \, , \\
     \Lambda_{ik}^{(b)}&=\,\frac{2\alpha m_f^2\,\mathcal{I}_{-2}(m_fr)}{3\pi^2r} (\hat{r}_i\hat{r}_k-\delta_{ik}) \, .
\end{align}
To first order in $\alpha$, the magnetic susceptibility is defined through the relation
\begin{equation}\label{chitensordef}
    \mathcal{M}_i^{(1)}=\chi_{ij}^{(1)} (\mathbf{B}^{(0)}_{md})_j \, ,
\end{equation}
where $ (\mathbf{B}^{(0)}_{md})_j$ is given in Eq.~(\ref{b0}). It can be written as
\begin{equation}\label{Xitensor}
   (\mathbf{B}^{(0)}_{md})_j=\Xi_{jl}m_l \,,
\end{equation}
with
\begin{equation}
    \Xi_{jl}=\frac{(3\hat{r}_j\hat{r}_l-\delta_{jl})}{4\pi r^3} \, .
\end{equation}
The magnetic susceptibility is readily obtained from Eqs.(\ref{chitensordef}) and (\ref{Xitensor}) and is given by $\chi^{(1)}=\chi^{(1a)}+\chi^{(1b)}$ with
\begin{eqnarray}
    \chi^{(1a)}_{ij}&=& \Lambda_{ik}^{(a)} \Xi^{-1}_{kj} \, ,\\
    \chi^{(1b)}_{ij}&=& \Lambda_{ik}^{(b)} \Xi^{-1}_{kj} \, .
\end{eqnarray}
The reader may readily verify that
\begin{equation}
    \Xi^{-1}_{kj} = 4\pi r^3\left(3\frac{\hat{r}_k\hat{r}_j}{2}-\delta_{kj}\right) \, ,
\end{equation}
leading us to
\begin{align}
    \chi^{(1a)}_{ij}&=\, \frac{\alpha}{6\pi^2r^3}\Big(\mathcal{I}_0(m_fr) + 2m_fr \mathcal{I}_{-1}(m_fr)\Big)\delta_{ij}=:\chi_a\delta_{ij} \, , \\
     \chi^{(1b)}_{ij}&=\, \frac{8\alpha }{3\pi} m_f^2r^2\,\mathcal{I}_{-2}(m_fr)(\delta_{ij}-\hat{r}_i\hat{r}_j)=:\chi_{b} (\delta_{ij}-\hat{r}_i\hat{r}_j)\, .
\end{align}
The second contribution shows that already at order $\alpha$ the vacuum exhibits an anisotropic magnetic response. Indeed, the magnetization is parallel to the magnetic only for  $\boldsymbol{r}\parallel\mathbf{m}$  or $\mathbf{r}\perp\mathbf{m}$. 

To assess the diamagnetic/paramagnetic character of the vacuum at order $\alpha$,  it suffices to compute the sign of $\boldsymbol{\mathcal{M}}^{(1)}\cdot\mathbf{B}^{(0)}_{md}$. Since $\chi_{a},\chi_b \geq 0$, we see that
\begin{align}
    \boldsymbol{\mathcal{M}}^{(1)}\cdot\mathbf{B}_{md}^{(0)}=\,\chi_a(&3(\mathbf{B}_{md}^{(0)}\cdot\mathbf{\hat{r}})^2+\mathbf{B}_{md}^{(0)2})\nonumber\\
    +&\chi_b(\mathbf{B}_{md}^{(0)2}-(\mathbf{\hat{r}}\cdot\mathbf{B}_{md}^{(0)})^2)\nonumber\\
      \geq 0 \, .
\end{align}
Therefore, the QED vacuum has a paramagnetic character. 

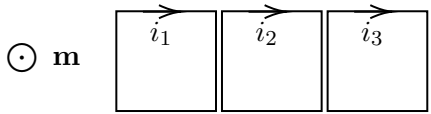
\begin{figure}[!h]
    \centering

\begin{tikzpicture}[x=0.75pt,y=0.75pt,yscale=-1,xscale=1]

\draw   (215,165) -- (265,165) -- (265,215) -- (215,215) -- cycle ;
\draw    (228.33,165.22) -- (245.33,165.22) ;
\draw [shift={(247.33,165.22)}, rotate = 180] [color={rgb, 255:red, 0; green, 0; blue, 0 }  ][line width=0.75]    (10.93,-3.29) .. controls (6.95,-1.4) and (3.31,-0.3) .. (0,0) .. controls (3.31,0.3) and (6.95,1.4) .. (10.93,3.29)   ;
\draw   (268,165) -- (318,165) -- (318,215) -- (268,215) -- cycle ;
\draw    (281.33,165.22) -- (298.33,165.22) ;
\draw [shift={(300.33,165.22)}, rotate = 180] [color={rgb, 255:red, 0; green, 0; blue, 0 }  ][line width=0.75]    (10.93,-3.29) .. controls (6.95,-1.4) and (3.31,-0.3) .. (0,0) .. controls (3.31,0.3) and (6.95,1.4) .. (10.93,3.29)   ;
\draw   (321,165) -- (371,165) -- (371,215) -- (321,215) -- cycle ;
\draw    (334.33,165.22) -- (351.33,165.22) ;
\draw [shift={(353.33,165.22)}, rotate = 180] [color={rgb, 255:red, 0; green, 0; blue, 0 }  ][line width=0.75]    (10.93,-3.29) .. controls (6.95,-1.4) and (3.31,-0.3) .. (0,0) .. controls (3.31,0.3) and (6.95,1.4) .. (10.93,3.29)   ;

\draw (158,179.54) node [anchor=north west][inner sep=0.75pt]    {$\bigodot \ \mathbf{m}$};
\draw (230.33,168.62) node [anchor=north west][inner sep=0.75pt]    {$i_{1}$};
\draw (283.33,168.62) node [anchor=north west][inner sep=0.75pt]    {$i_{2}$};
\draw (336.33,168.62) node [anchor=north west][inner sep=0.75pt]    {$i_{3}$};

\end{tikzpicture}

    \caption{Qualitative picture of the magnetization field $\boldsymbol{\mathcal{M}}$. Each current loop (of infinitesimal size) represents the magnetization at a given point. Since $|\boldsymbol{\mathcal{M}}|$ decreases with the distance from the point dipole, the current intensities satisfy $i_1>i_2>i_3>0$, which is consistent with the current density lines in Fig.~\ref{Fig:j}.  }
    \label{Fig:M}
\end{figure}

In condensed matter physics, the paramagnetic response of a material arises due to the spin of the constituting particles. Here we attribute the vacuum paramagnetic character to the spins of the fermionic particle-anti-particle continuous creation and annihilation. Note that the induced magnetization  is qualitatively consistent with the current density represented in Fig.~\ref{Fig:j}. Figure~\ref{Fig:M} illustrates this point. The question of the magnetic character of the vacuum has been explored in~\cite{rojas2,rojasPhysRevD.79.093002} and has also been investigated in the context of confinement in Yang--Mills theories in~\cite{PAGELS1978485}.

\section{Uehling Magnetic corrections in neutron-neutron scattering and hyperfine splitting}\label{section:applications}

 {In this section, we will discuss two scenarios in which the specific magnetic vacuum corrections obtained in the present work leave imprints in physical observables. We begin with a discussion of the effect in neutron-neutron interactions, which provide a clear regime where the Uehling magnetic correction becomes important. Then, we revisit the standard hyperfine splitting and discuss how the same correction also changes the specific form of the Landé interval factor.}

\subsection{Effective interaction with a neutral magnetic probe}
\label{section:Fabricio}

Here we analyze the long-distance neutron--neutron scattering. Although electrically neutral, the neutron has a nonvanishing electromagnetic current and a permanent magnetic moment originating from its charged quark constituents \cite{PunjabiEtAl2015,MohrEtAl2025}. At momentum transfers too small to resolve its internal structure, it can thus be represented as a pointlike magnetic dipole, with the hadronic dynamics encoded in the measured value of \(\mu_n\).

This setting is particularly interesting because the monopolar Coulomb interaction is absent, the strong nuclear interaction is suppressed at sufficiently large separations and, as we demonstrate below, the noncontact Maxwellian dipole--dipole interaction vanishes in the \({}^{1}\mathsf{S}_0\) channel. At distances not much larger than the electron Compton scale, the vacuum-polarization correction may therefore provide a distinct long-distance contribution. Our purpose is not to construct a complete neutron--neutron interaction or to claim dominance over all residual effects, but to provide the universal point-dipole kernel that can enter such a description.
At low relative momenta, neutron--neutron scattering is dominated by the lowest partial wave allowed by the exchange symmetry of the two-neutron state. Since neutrons are identical spin-\(1/2\) fermions, their complete two-particle state must be antisymmetric under particle exchange. For an \(\mathsf{S}\) wave, \(L=0\), the spatial wave function is symmetric. The two-neutron isospin state is also symmetric, with \(T=1\), and the spin state must therefore be antisymmetric, corresponding to total spin \(S=0\). This channel is conventionally denoted by
\begin{equation}
  {}^{2S+1}L_J = {}^{1}\mathsf{S}_0,
\end{equation}
where the superscript gives the spin multiplicity \(2S+1\), the letter
\(\mathsf{S}\) denotes orbital angular momentum \(L=0\), and the subscript is
the total angular momentum \(J=0\). The \({}^{1}\mathsf{S}_0\) neutron--neutron
channel plays a central role in low-energy neutron physics
\cite{Sammarruca2023}. 

Treating the two-neutron system as a pair of point-like magnetic dipoles, we write their magnetic dipole moment operators as
\begin{equation}
  \boldsymbol{\mu}_a
  =
  \mu_n\boldsymbol{\sigma}_a,
  \qquad
  a=1,2,
\end{equation}
where $\mu_n$ denotes the intrinsic magnetic dipole moment of the neutron and the Pauli matrices \(\boldsymbol{\sigma}_a\) act on the spin of neutron \(a\).
Away from the position of the source, the Maxwellian dipole field produces the interaction
\begin{equation}
V_{\mathrm{dd}}^{(0)}(\boldsymbol{r})
=
-\boldsymbol{\mu}_2\cdot
\boldsymbol{B}^{(0)}_1(\boldsymbol{r})
= -
\frac{1}{4\pi r^3} \mu_n S_{12},
\qquad r>0, 
\label{eq:classical-nn-magnetic}
\end{equation}
where $S_{12} \equiv
3
\bigl(\boldsymbol{\sigma}_1\cdot\hat{\boldsymbol{r}}\bigr)
\bigl(\boldsymbol{\sigma}_2\cdot\hat{\boldsymbol{r}}\bigr)
-
\boldsymbol{\sigma}_1\cdot\boldsymbol{\sigma}_2$. 
Magnetic-moment interactions of this type are included in standard descriptions of the electromagnetic sector of the nucleon--nucleon interaction \cite{WiringaEtAl1995}. 

In the spin-singlet state,  one has
\begin{equation}
\left\langle
\sigma_{1i}\sigma_{2j}
\right\rangle_{S=0}
=
-\delta_{ij},
\label{eq:singlet-spin-correlation}
\end{equation}
and hence
\begin{equation}
\begin{split}
\left\langle
\boldsymbol{\sigma}_1\cdot\boldsymbol{\sigma}_2
\right\rangle_{S=0}
&=
-3,
\\
\left\langle
\bigl(\boldsymbol{\sigma}_1\cdot\hat{\boldsymbol{r}}\bigr)
\bigl(\boldsymbol{\sigma}_2\cdot\hat{\boldsymbol{r}}\bigr)
\right\rangle_{S=0}
&=
-1.
\end{split}
\label{eq:singlet-contractions}
\end{equation}
Consequently, the two contributions in Eq.~\eqref{eq:classical-nn-magnetic} coming form $S_{12}$ cancel exactly,
\begin{equation}
\begin{split}
\left\langle
V_{\mathrm{dd}}^{(0)}(\boldsymbol{r})
\right\rangle_{{}^{1}\mathsf{S}_0}
&= 0,
\qquad r>0.
\end{split}
\label{eq:classical-singlet-cancellation}
\end{equation}
 This
cancellation is not a consequence of the absence of magnetic moments,
but of the precise relative factor \(3\) between the two tensor
structures of the Maxwellian dipole field.

The vacuum-polarization correction derived in the present work has the general tensorial structure
\begin{equation}
\delta\boldsymbol{B}_1(\boldsymbol{r})
=
f^{(1)}_1(r)\boldsymbol{\mu}_1
+
f^{(1)}_2(r)
\bigl(
\boldsymbol{\mu}_1\cdot\hat{\boldsymbol{r}}
\bigr)
\hat{\boldsymbol{r}},
\label{eq:corrected-magnetic-field-general}
\end{equation}
 where \(f^{(1)}_1(r)\) and \(f^{(1)}_2(r)\), which are the radial functions generated by
the vacuum-polarization insertion, can be read off from Eq.~\eqref{B_field_correction} and are explicitly given by
\begin{align}
    f^{(1)}_1(r) &= -\frac{\alpha}{6\pi^2 r^3}(\mathcal I_0 +2m_f r \mathcal I_{-1} + 4m_f^2 r^2 \mathcal I_{-2})\,,\\
    f^{(1)}_2(r) &= \frac{\alpha}{6\pi^2 r^3}(3\mathcal I_0 + 6 m_f r \mathcal I_{-1} + 4m_f^2 r^2 \mathcal I_{-2})\,.
\end{align}

The corresponding correction to the interaction with a second point-like neutron is
\begin{equation}
\begin{split}
\delta V_{12}(\boldsymbol{r})
&=
-\boldsymbol{\mu}_2\cdot
\delta\boldsymbol{B}_1(\boldsymbol{r})
\\
&=
-f^{(1)}_1(r)
\boldsymbol{\mu}_1\cdot\boldsymbol{\mu}_2
\\
&\quad
-f^{(1)}_2(r)
\bigl(
\boldsymbol{\mu}_1\cdot\hat{\boldsymbol{r}}
\bigr)
\bigl(
\boldsymbol{\mu}_2\cdot\hat{\boldsymbol{r}}
\bigr).
\end{split}
\label{eq:corrected-nn-interaction}
\end{equation}
Using
\(\boldsymbol{\mu}_a=\mu_n\boldsymbol{\sigma}_a\) and the definition of \(S_{12}\), Eq.~\eqref{eq:corrected-nn-interaction} can equivalently be written as
\begin{equation}
\begin{split}
\delta V_{12}(\boldsymbol{r})
=
-\mu_n^2
\biggl[
&
\left(
f^{(1)}_1(r)+\frac{f^{(1)}_2(r)}{3}
\right)
\boldsymbol{\sigma}_1\cdot\boldsymbol{\sigma}_2
\\
&+
\frac{f^{(1)}_2(r)}{3}S_{12}
\biggr].
\end{split}
\label{eq:corrected-nn-central-tensor}
\end{equation}
This decomposition separates a central spin--spin contribution from the tensor contribution. In the \({}^{1}\mathsf{S}_0\) channel,
\(\langle S_{12}\rangle_{{}^{1}\mathsf{S}_0}=0\) and
\(\langle\boldsymbol{\sigma}_1\cdot\boldsymbol{\sigma}_2
\rangle_{{}^{1}\mathsf{S}_0}=-3\). Consequently,
\begin{equation}
\left\langle
\delta V_{12}(\boldsymbol{r})
\right\rangle_{{}^{1}\mathsf{S}_0}
=
\mu_n^2
\left[
3f^{(1)}_1(r)+f^{(1)}_2(r)
\right].
\label{eq:uehling-singlet-interaction}
\end{equation}
For comparison, the Maxwellian dipole field corresponds, for
\(r>0\), to
\begin{equation}
f_1^{(0)}(r)
=
-\frac{1}{4\pi r^3},
\qquad
f_2^{(0)}(r)
=
\frac{3}{4\pi r^3},
\label{eq:maxwell-radial-functions}
\end{equation}
and therefore satisfies
\begin{equation}
3f_1^{(0)}(r)+f_2^{(0)}(r)=0.
\label{eq:maxwell-radial-cancellation}
\end{equation}

{The cancellation of the classical \({}^{1}\mathsf{S}_0\) matrix element is thus a consequence of the specific relation \(f_2^{(0)}=-3f_1^{(0)}\), which is not satisfied by the $f^{(1)}_1, f^{(1)}_2$ radial functions.} As a result, the noncontact vacuum-polarization correction need not vanish in the spin-singlet channel. 
It is therefore the first nonvanishing contribution of this type to the long-distance \({}^{1}\mathsf{S}_0\) interaction.

The separation of scales makes this observation particularly natural for two neutrons at large relative distances. The longest-range part of the strong nucleon--nucleon interaction is governed by one-pion exchange, whose asymptotic behavior is controlled by \(\exp(-m_\pi r)\) \cite{PiarulliTews2020}. By contrast, the electronic vacuum-polarization correction is controlled asymptotically
by \(\exp(-2m_e r)\). Since
\begin{equation}
m_\pi^{-1}\simeq 1.4~\mathrm{fm},
\qquad
(2m_e)^{-1}\simeq 193~\mathrm{fm},
\end{equation}
there is a parametrically separated range
\begin{equation}
m_\pi^{-1}\ll r\lesssim (2m_e)^{-1},
\label{eq:nn-distance-window}
\end{equation}
in which the direct strong-interaction tail is exponentially suppressed while the electronic vacuum-polarization correction has not yet reached its far asymptotic suppression. These distances are also much larger than the internal size of the neutron, consistently with the point-dipole approximation adopted here.

The point-like approximation applies to the electromagnetic vertices representing the neutron magnetic moments; it does not require the
relative neutron state to be spatially localized. In a scattering
problem, Eq.~\eqref{eq:corrected-nn-interaction} can instead be used
as a long-distance interaction kernel acting on the delocalized
two-neutron scattering state. A quantitative calculation of phase shifts or scattering observables would require embedding this kernel in a complete nuclear interaction model, including the strong interaction and other electromagnetic contributions, and evaluating it with the appropriate scattering states. Such an analysis lies beyond the scope of the present field-theoretical treatment. Our more limited observation is that the corrected point-dipole field generates a well-defined noncontact contribution that survives projection onto the \({}^{1}\mathsf{S}_0\) channel, whereas the corresponding Maxwellian contribution vanishes.

\subsection{Application to Hyperfine Structure \label{hyperfine_appendix}}

As discussed in the previous subsection, the imprint of the Uehling vacuum polarization is manifest in neutron interactions under a certain regime. Here we discuss how the corrections also leave signatures in the more well-known case of the hyperfine splitting. 

The hyperfine structure is a very important set of corrections to atomic spectra that take into consideration the interaction of electrons with the nucleus. In the literature, the first correct assessment of this phenomenon from a theoretical point of view was given by Fermi~\cite{fermi1930magnetic, fermi1933theorie} and was subsequently analyzed by various authors~\cite{rosenthalPhysRev.41.459,arnowitt1953hyperfine,weiskopfPhysRev.77.94,naferabiPhysRev.71.914}. The hyperfine correction has been particularly important in the analysis of radiation from intergalactic hydrogen---leading to the famous 21 cm shift~\cite{pritchard201221,furlanetto200621,horii2017can,scott199021}---and helium~\cite{mcquinn2009redshifted} (see also~\cite{purcell1956influence} for an account of the effect in hydrogen atomic collisions). In this  appendix, we consider the problem of finding the Uehling-like vacuum correction to the hyperfine structure (for vacuum corrections including the Wichmann-Kroll potential, following a different approach,  see \cite{Artemyev2001}). Specifically, we evaluate the leading correction due to vacuum magnetization to the hyperfine structure directly from the electromagnetic fields obtained in this paper.

Pragmatically, we wish to compute the corrections arising from the potential and magnetic field corrections $\delta\vb{A}^{(1)},\, \delta\vb{B}^{(1)}$ written above, to the energy levels of a particle of angular momentum $\vb{J} = \vb{L}+\vb{S}$ moving in the field of a nucleus with angular momentum $\vb{I}$. This calculation corresponds to using perturbation theory on the non-relativistic Hamiltonian for a spin 1/2 particle bounded to the nucleus, i.e.
\begin{eqnarray}
    H = H_0 + V_C + H_{FS} + H_{Lamb} + H_{hf}\,,
\end{eqnarray}
with the hyperfine term $H_{hf}$ given by
\begin{eqnarray}
    H_{hf} = - \vb{p}\cdot \vb{A}^{(1)} - {\vb*\mu}\cdot\vb{B}^{(1)}\,,
\end{eqnarray}
where $\vb*{\mu} = -\mu_e g_e\vb{S}$ is the electron's magnetic moment, $\mu_e = e/2m_e$ and $g_e$ is the gyromagnetic factor, $\vb{p}$ is the electrons momentum, $\vb{A}^{(1)}$ is the potential generated by the nucleus, $V_C$ is the central (Coulomb) potential and the other terms correspond to the corrections due to the fine structure, Lamb shift and the hyperfine structure, respectively. Of all of these terms, the hyperfine corrections are the smallest, which motivates the use of perturbation theory to calculate its effects.

Now, the effect of the other terms in the Hamiltonian is to create an explicit dependence of the energy levels of the electron on the electronic quantum numbers $n,j,l$.
 By including the nucleus, with its own angular momentum $I$, we consider a Hilbert space composed of the direct product between the electron's and the nucleus's spaces, that is, $\mathcal{E} = \mathcal{E}_e \otimes \mathcal{E}_{N}$. A possible basis for these states is simply $\ket{nljm_j}\otimes \ket{I M_I}$, with $-j\le m_j\leq j$, $-I\leq M_I\leq I$ representing the spin projection numbers. A better basis, which enables the diagonalization of the degenerate subspace, is the one constructed with the sum of the angular momenta of the nucleus and the electron, $\vb{F} = \vb{J} + \vb{I}$, represented by states $\ket{nlj I FM_F}$. For the first order perturbation, then, we would have to evaluate
\begin{eqnarray}
    \Delta E = \bra{nljI FM_F} H_{hf} \ket{nljIFM_F}\,.
\end{eqnarray}

In the standard case, we would proceed to write the field $\vb{B}$ and the potential $\vb{A}$ generated by the nucleus and plug them in the above formula. In our case, however, we are considering $\vb{A}^{(1)}$ and $\vb{B}^{(1)}$, that is, the leading $\alpha$ corrections to these expressions. These are given in eq.~\eqref{dipole_potential1} and eq.~\eqref{Magnetic_dipole_field}, and in this context can be written as
\begin{eqnarray}
    \vb{A}^{(1)} &=& \mu_Ng_N\frac{\vb{I}\times \vb{r}}{4\pi r^3}\left(1 + \frac{2\alpha}{3\pi} \Phi_{m_f}(r)\right) ,\\
    \vb{B}^{(1)} &=& \mu_N g_N\frac{3 \vb{r} (\vb{I}\cdot \vb{r}) - r^2\vb{I}}{4\pi r^5}\left(1 + \frac{2\alpha}{3\pi} \Phi_{mf}(r)\right) \nonumber\\
    &&+ \mu_N g_N\frac{\alpha}{6\pi^2} \frac{\vb{r}(\vb{I}\cdot\vb{r}) - r^2\vb{I}}{r^5}\Psi_{m_f}(r)\nonumber\\
    &&+ \frac{2}{3}\mu_N g_N \delta(\vb{r})\,\vb{I}\left(1 + \frac{2\alpha}{3\pi} \Phi_{mf}(r)\right),  \label{eq.B(1)appendix}
\end{eqnarray}
where $\Phi_{mf} = \mathcal{I}_0(m_fr) + 2m_f r \mathcal{I}_{-1}(m_fr)$ and $\Psi_{mf} = 4m_f^2 r^2 \mathcal{I}_{-2}(m_fr)$. By using some of the general properties of angular momentum operators, the above can be evaluated in terms of the expectation values of the radial components. These values depend on the specific wave function one considers for the particle. The outcome of the analysis is that the full correction would take the form
\begin{eqnarray}
    \Delta E = \Delta E^{(\alpha^0)} + \Delta E^{(\alpha^1)} 
\end{eqnarray}
where the corrections to the ground state, for $l=0$, are
\begin{eqnarray}
    \Delta E_{l=0}=
     \frac{1}{3}g_e g_N \mu_e \mu_N \big(F(F+1) - I(I+1) - 3/4\big) \nonumber\\
     \times\expval{\delta(\vb{r})\left( 1 + \frac{2\alpha}{3\pi} \Phi_{mf}(r)\right)}\,,\label{hyper:ground_state_corr}
\end{eqnarray}
and the corrections to the $l \neq 0$ states, which we separate in the $\alpha^0$ and $\alpha^1$ orders, are given by
\begin{eqnarray}
    \Delta E_{l\neq 0}^{(\alpha^0)} &=& \frac{g_e g_N \mu_e \mu_N}{4\pi} \frac{F(F+1) - j(j+1) - I(I+1)}{2j(j+1)} \nonumber\\
    &&\times l(l+1) \expval{\frac{1}{r^3}}\,,\label{hyper:0ordercorrection}\\
    \Delta E_{l\neq 0}^{(\alpha^1)} &=&\frac{\alpha g_e g_N \mu_e\mu_N}{6\pi^2}\frac{F(F+1) - j(j+1) - I(I+1)}{2j(j+1)}\nonumber\\
    &&\times\Big(l(l+1) \expval{\frac{\Phi_{mf}(r)}{r^3}}\nonumber\\ &&+ \frac{1}{2}\left(l(l+1) - j(j+1) - \frac{1}{4}\right)\expval{\frac{\Psi_{mf}(r)}{r}}\Big)\,.\nonumber\\ \label{hyper:alphaordercorrection}
\end{eqnarray}

As mentioned above, the expectation values are dependent on the specific wavefunction one considers. Furthermore, the above calculation considers a single fermion in the loop, while the full contribution would be a sum of the above $\Delta E^{(\alpha^1)}$ correction over the masses of all the fermions  in the theory.  Interestingly, we see that the   angular dependence coming from the second term in the right-hand side of Eq.~(\ref{eq.B(1)appendix})   leads to a modification  of the electron quantum number dependence, which corresponds to a deviation from the standard Land\'e interval rule. 

It is important to note that, as in the case of the point particle, the vacuum polarization contribution to atomic spectra is not the full story and yields the smallest contribution to the Lamb shift. In the same spirit, if one attempts to perform the calculation of the energy shifts for the $s$ states, the point particle approximation for the nucleus must of course be taken with a grain of salt.

A quick illustration of this comes from using the small $\vb{p}^2/m_f^2$  approximation of the vacuum polarization tensor. For $\vb{p}^2/m_f^2<<1$, the polarization tensor is approximated as
\begin{eqnarray}
    \Pi_R(-\vb{p}^2) &=& -\frac{2\alpha}{\pi} \sum_{n=1}^{\infty}\frac{(-1)^n}{n}\left(\frac{\vb{p}^2}{m_f^2}\right)^n \frac{[(n+1)!]^2}{(2n+3)!}\nonumber\\
    &\approx& \frac{\alpha}{\pi} \frac{\vb{p}^2}{15m_f^2} + O(\vb{p}^2/m^2)\,,
\end{eqnarray}
such that, after taking $\epsilon \to 0$, one finds
\begin{eqnarray}
    \int\frac{d^3\vb{p}}{(2\pi)^3} \frac{\Pi_R(-\vb{p}^2)}{-\vb{p}^2 +i\epsilon} e^{i\vb{p}\cdot \vb{r}} \approx -\frac{\alpha}{15m_f^2\pi} \delta(\vb{r})
\end{eqnarray}  
Using the approximation described above to leading order in $\vb{p}^2/m^2_f$, the functions $\Phi_{mf}, \Psi_{mf}$ can be written in terms of delta functions at the origin. For the $s$ states, only the former appear and one has 
\begin{eqnarray}
    \Phi_{mf}(r) \approx \frac{6\pi r}{5m_f^2}\delta(\vb{r})\,.
\end{eqnarray}
With this, the correction to the energy levels would be proportional to $\expval{\delta(\vb{r}) (1 + 4\alpha r \delta(\vb{r})/5m_f^2}$.
Then, once the expectation value is calculated for a standard $s$ orbital Hydrogen-like wavefunction, one finds that the $\alpha$ correction vanishes in the distributional sense. Of course, as mentioned, this is an artifact of the point particle approximation of the nucleus itself which the low $\vb{p}^2/m_f^2$ approximation of the vacuum polarization cloud probes. This same problem can be seen even without carrying out the approximation, in the form of a logarithmic divergence which comes from the convolution of the $\Phi_{mf}$ integral with the $\delta(\vb{r})$ from the Fermi contact term.

\section{Conclusions}\label{secion:conclusion}

In this work, we have calculated the generalization of the original Uehling correction to a point magnetic dipole field. We also computed the Uehling effect for an electric point dipole in order to compare it with the magnetic case. We demonstrated that the symmetry present in classical electromagnetism relating the electric field produced by a point electric dipole and the magnetic field generated by a point magnetic dipole is broken due to the loss of scale invariance entailed by the presence of the fermionic mass.

Furthermore, by considering the induced vacuum currents, we found properties which correspond to the paramagnetic character of the QED vacuum at linear order in $\alpha$. Indeed, already at this order we see that magnetic susceptibility presents a nontrivial tensorial structure. Other results reported in the literature for the paramagnetic behavior of the QED vacuum account for an anisotropic response only at second order and in dynamical situations, as for example in the phenomenon of vacuum birefringence in the presence of an external field~\cite{GiesPhysRevD.92.071301,zavattini,bragin2017high}.

As noted in the text, the interpretation of these quantum properties in terms of the condensed matter point of view might not only furnish a physical picture for the highly non-intuitive  behavior of the quantum vacuum, but  also provide ways of extracting important information concerning the general properties of a given theory,  e.g. in the case of confinement~\cite{PAGELS1978485}. Similarly, as we exemplified with the simple hydrogen-like toy model, the corrected fields provide a simple way to estimate the leading correction to the hyperfine structure.

We also note that the behavior of the renormalized coupling at the level of the fields seems to indicate that the separation of “dipole” and   “monopole” fields within the quantum corrections is not as trivial as one might expect. This is because the dipole corrections do not correspond to a simple renormalization of the dipole moments, i.e. the radial dependence of the renormalized coupling gives rise to more terms which do not fit into the tensor structures associated with the standard multipole operators. It would be interesting to see if this property carries on to higher multipole orders and if one could extract an underlying pattern for the corrections. At any rate, it is precisely because of this nontrivial dipole correction that the  magnetic Uehling correction leaves an important signature in the long-distance neutron-neutron interaction, as we have shown. 

 {We expect our results to provide a useful framework for investigating vacuum-polarization effects in dipolar configurations. In particular, the corrected magnetic field may serve as a building block for more complete studies of low-energy nuclear and muonic systems whose electromagnetic dynamics is influenced by permanent nuclear magnetic dipole moments.}

\begin{acknowledgments}

  The work of G.Z. was partly supported by CAPES and CNPQ. C.F. acknowledges funding from CNPq (Grants No. 308641/2022-1 and 408735/2023-6) and FAPERJ (Grant No. 204.376/2024). F.A.B thanks CNPq under the grant 313426/2021-0.

\end{acknowledgments}

\appendix

\section{Dipole symmetry within Maxwell electrodynamics \label{appendix:dipole_symmetry}}

As mentioned in the text there is a symmetry between the fields produced by an electric dipole and by a magnetic dipole in Maxwell electrodynamics. More precisely, denoting the fields produced by an electric (magnetic) point dipole with a subscript ed (md) the symmetry is given by the mapping $\mathbf{d}\to\mathbf{m}$, $\mathbf{B}_{\rm ed}\to-\mathbf{E}_{\rm md}$ and $\mathbf{E}_{\rm ed}\to \mathbf{H}_{\rm md}=\mathbf{B}_{\rm md}+\mathbf{m}\delta(\boldsymbol{r})$, where $\mathbf{H}$ denotes the induction magnetic field and $\mathbf{m}\delta(\mathbf{r})$ is the magnetization density.  From the perspective of the sources involved this is a surprising result, since the structure of the charge density and the current density is very different for the two cases. Here we present an elementary derivation of this symmetry, without the need of actually evaluating the fields. We consider the general dynamic situation, where the point dipoles can be time dependent. The charge and current distributions corresponding to an electric dipole are given by~\cite{pitombo2021}
\begin{eqnarray}
    \rho_{\rm ed}(\mathbf{r},t)&=&-\mathbf{d}(t)\cdot\boldsymbol{\nabla}\delta(\mathbf{r}) \, ,\\
       \mathbf{J}_{\rm ed}(\mathbf{r},t)&=&\dot{\mathbf{d}}(t)\delta(\mathbf{r}) \, .
\end{eqnarray}
The sources corresponding to a magnetic dipole, on the other hand, are given by
\begin{eqnarray}
    \rho_{\rm md}(\mathbf{r},t)&=&0\, ,\\
       \mathbf{J}_{\rm md}(\mathbf{r},t)&=&-\mathbf{m}(t)\times\boldsymbol{\nabla}\delta(\mathbf{r}) \, .
\end{eqnarray}
In Maxwell electrodynamics, the fields produced can be obtained by solving the inhomogeneous wave equation given by
\begin{align}
    \left(\boldsymbol{\nabla}^2-\frac{1}{c^2}\frac{\partial^2}{\partial t^2}\right)\mathbf{E} &=\, 4\pi\boldsymbol{\nabla}\rho+\frac{4\pi}{c^2}\mathbf{\dot{J}} \, ,\label{ewave}\\ 
      \left(\boldsymbol{\nabla}^2-\frac{1}{c^2}\frac{\partial^2}{\partial t^2}\right)\mathbf{B} &=\, -\frac{4\pi}{c^2}\boldsymbol{\nabla}\times\mathbf{J}  \, . \label{bwave}
\end{align}
Note that $\boldsymbol{\nabla}\times\boldsymbol{J}_{\rm ed}=-\mathbf{\dot{d}}\times\boldsymbol{\nabla}\delta(\mathbf{r})$ and $\mathbf{\dot{J}}_{\rm md}=-\mathbf{\dot{m}}\times\boldsymbol{\nabla}\delta(\mathbf{r})$. This shows that under the transformation $\boldsymbol{d}\to \boldsymbol{m}$, the magnetic field $\mathbf{B}_{\rm ed}$ produced by the electric dipole is taken into $-\mathbf{E}_{\rm md}$,
thus establishing the first part of the symmetry. The other pair of fields is more subtle, so we write the differential equations explicitly:
\begin{align}
      \left(\boldsymbol{\nabla}^2-\frac{1}{c^2}\frac{\partial^2}{\partial t^2}\right)\mathbf{E}_{\rm ed} &=\, -4\pi\left((\mathbf{d}\cdot\boldsymbol{\nabla})\boldsymbol{\nabla}-\frac{1}{c^2}\frac{\partial^2\mathbf{d}}{\partial t^2}\right)\delta(\mathbf{r})
      \, ,\label{edede}\\
      \left(\boldsymbol{\nabla}^2-\frac{1}{c^2}\frac{\partial^2}{\partial t^2}\right)\mathbf{B}_{\rm md} &=\, -4\pi\left((\mathbf{m}\cdot\boldsymbol{\nabla})\boldsymbol{\nabla}-\boldsymbol{\nabla}^2\mathbf{m}\right)\delta(\mathbf{r}) \, .
\end{align}
By subtracting $(\boldsymbol{\nabla}^2-\frac{1}{c^2}\partial_t^2)\mathbf{m}\delta(\mathbf{r})$ on both sides of the last equation we can recast it in the form 
\begin{equation}
    \left(\boldsymbol{\nabla}^2-\frac{1}{c^2}\frac{\partial^2}{\partial t^2}\right)\mathbf{H}_{\rm md}=-4\pi\left((\mathbf{m}\cdot\boldsymbol{\nabla})\boldsymbol{\nabla}-\frac{\partial^2\mathbf{m}}{\partial t^2}\right)\delta(\mathbf{r}) \, .
\end{equation}
Comparing with equation (\ref{edede}) we see that under the transformation $\mathbf{d}\to\mathbf{m}$ we have the mapping $\mathbf{E}_{\rm ed}\to\mathbf{H}_{\rm md}$. For $\mathbf{r}\neq \mathbf{0}$ we have $\mathbf{H}_{\rm md}=\mathbf{B}_{\rm md}$ and thus we can interpret our results by establishing that a magnetic loop in the dipole approximation produces the same field as two fictitious monopole magnet moments with opposite magnetic charge. However, these sources are not equivalent when we need the field at the origin~\cite{griffiths1982hyperfine,azevedo2024can}.

\section{Calculation steps from \eqref{eq:3.16} to \eqref{eq:3.17} \label{apendix1}}

Using the explicit renormalized polarization tensor, the second integral with $p_i p_j$ takes the form
\begin{eqnarray}
    \partial_i \partial_j\int\frac{\dd[3]{\vb{p}}}{(2\pi)^3} \frac{1}{\vb{p}^4} e^{i\vb{p}\cdot \Delta\vb{x}} \frac{\alpha}{\pi}\frac{\vb{p}^2}{4m_f^2} \int_0^1 \dd{v} \frac{v^2(1-\frac{1}{3}v^2)}{1 + (\vb{p}/2m_f)^2[1-v^2]}\nonumber\,,
\end{eqnarray}
or, exchanging the order of integrations,
\begin{eqnarray}
    &&\frac{\alpha}{4m_f^2\pi} \partial_i \partial_j \int_0^1 \dd{v} v^2(1-\frac{1}{3}v^2) \nonumber\\
    &&\times\int\frac{\dd[3]{\vb{p}}}{(2\pi)^3} \frac{1}{\vb{p}^2} \frac{1}{1 + (\vb{p}/2m_f)^2[1-v^2]} e^{i\vb{p}\cdot \Delta\vb{x}}\,.
\end{eqnarray}
For now, let us focus on the $\vb{p}$ integration. In spherical polar coordinates with $\dd[3]{\vb{p}} = p^2 \dd{p} \dd\cos\theta \dd\varphi$, we write the radial variable as $p = |\vb{p}|(1-v^2)^{1/2}/2m_f$ such that the above becomes
\begin{eqnarray}
     &&\int\frac{\dd[3]{\vb{p}}}{(2\pi)^3} \frac{1}{\vb{p}^2} \frac{1}{1 + (\vb{p}/2m_f)^2[1-v^2]} e^{i\vb{p}\cdot \Delta\vb{x}}\nonumber\\
     &&= \frac{1}{2\pi^2 |\Delta\vb{x}|} \int_0^\infty \frac{\dd{p}}{p(1+p^2)} \sin\left(\frac{2m_f|\Delta\vb{x}|}{\sqrt{1 - v^2}}p\right) \,.\nonumber
\end{eqnarray}
The above integral can be evaluated using contour integration:
\begin{eqnarray}
    \int_0^\infty \frac{\dd{p}}{p(1+p^2)} \sin\left(\Omega p\right) &=& \frac{1}{2}\Im \int_{-\infty}^\infty \frac{\dd{p}}{p(1+p^2)}  e^{i\Omega p}\nonumber\\
    &=& \frac{\pi}{2}(1 - e^{-\Omega})\,.
\end{eqnarray}
Therefore, we get:
\begin{eqnarray}
    &&\int\frac{\dd[3]{\vb{p}}}{(2\pi)^3} \frac{1}{\vb{p}^2} \frac{1}{1 + (\vb{p}/2m_f)^2[1-v^2]} e^{i\vb{p}\cdot \Delta\vb{x}}\nonumber\\
    &=&\frac{1}{2\pi^2 |\Delta\vb{x}|}\frac{\pi}{2}\left[1 - \exp\left(-\frac{2m_f|\Delta\vb{x}|}{\sqrt{1 - v^2}}\right) \right]\,,
\end{eqnarray}
and, thus, we have
\begin{eqnarray}
     &&\frac{\alpha}{4m_f^2\pi} \partial_i \partial_j \frac{1}{4\pi |\Delta\vb{x}|}\int_0^1 \dd{v} v^2(1-\frac{1}{3}v^2)\nonumber\\
     &&\times\left[1 - \exp\left(-\frac{2m_f|\Delta\vb{x}|}{\sqrt{1 - v^2}}\right) \right]\,.
\end{eqnarray}
Now, we can reexpress the above in terms of $\xi = (1-v^2)^{-1/2}$. Note, however, that the first piece (the one in square brackets) is immediately integrated in $v$, giving $4/15$. We can, thus, write
\begin{eqnarray}
     &&\frac{\alpha}{4m_f^2\pi} \partial_i \partial_j\Big\{ \frac{1}{4\pi |\Delta\vb{x}|} \nonumber\\
     &&\times\Big[\frac{4}{15}- \frac{2}{3}\int_1^\infty \dd{\xi} \frac{\sqrt{\xi^2 - 1}}{\xi^4}\left(1 + \frac{1}{2\xi^2}\right)e^{-2m_f|\Delta\vb{x}|\xi}\Big]\Big\}\,.\nonumber\\
\end{eqnarray}
Application of the derivatives is straightforward, albeit tedious. For the sake of notation, we take $\Delta\vb{x} = \vb{r} = \vb{x}-\vb{x}_s$, such that $|\Delta\vb{x}| = r$. Then, the above becomes
\begin{align}
    &\int\frac{\dd[3]{\vb{p}}}{(2\pi)^3} \frac{\Pi_R(-\vb{p}^2)}{-\vb{p}^4} p_i p_j e^{i\vb{p}\cdot \Delta\vb{x}} \nonumber\\
    &=\frac{\alpha}{16m_f^2 \pi^2}\Big\{\frac{4}{15r^3}\left(\frac{3 r_i r_j}{r^2} - \delta_{ij} \right)+\frac{2}{3}\int_1^\infty \dd{\xi} \frac{U(\xi)}{\xi^2} \frac{e^{-2m_f\xi r}}{r^2}\nonumber\\
    &\times\left.\left[ -4m_f^2 \xi^2 \frac{r_i r_j}{r} + \frac{1}{r}\left( \delta_{ij} - \frac{3r_i r_j}{r^2}\right)\left(1 + 2m_f\xi r \right)\right]\right\}.\label{magdipole:almost}
\end{align}

Now, we recall the first piece of the projector. From the standard Uehling correction, this is simply
\begin{eqnarray}
    \int\frac{\dd[3]{\vb{p}}}{(2\pi)^3} \frac{\Pi_R(-\vb{p}^2)}{-\vb{p}^2 + i\epsilon} \delta_{ij} e^{i\vb{p}\cdot(\vb{x} - \vb{x}_s)}\nonumber\\= -\frac{\alpha}{6\pi^2r}\delta_{ij} \int_1^\infty \dd{\xi} U(\xi) e^{-2m_f\xi r}\,.\label{magdipole:simple}
\end{eqnarray}
Combining Eqs.~\eqref{magdipole:almost} and ~\eqref{magdipole:simple}, we arrive at \eqref{eq:3.17}.

\section{Analytic Expressions\label{analytic}}

The field corrections calculated in Eqs.~\eqref{E_field_correction},~\eqref{B_field_correction} can be expressed in terms of Bickley-Naylor functions, defined as iterated integrals of modified Bessel functions. The fact that these arise in the integrals related to the Uehling correction is well-known in the literature~\cite{painjean,frolov2012}, but their use also appears in other contexts~\cite{bickley_naylorgrav2}. In general, the integrals $\mathcal{I}_n(z)$ as defined in the text can be expressed as
\begin{eqnarray}
    \mathcal{I}_n(z) &:=& \int_1^\infty d\xi\ U(\xi)\xi^{-n} e^{-2 z \xi}\nonumber\\&=& \text{Ki}_{n}(2z) - \frac{1}{2}\left[\text{Ki}_{n+2}(2z) + \text{Ki}_{n+4}(2z) \right]
\end{eqnarray}
where the functions $\text{Ki}_n(z)$ are defined as
\begin{eqnarray}
    \text{Ki}_n(z)&:=& \int_z^\infty K_{n-1}(x) dx \nonumber\\&=& \int_0^\infty \frac{e^{-z \cosh(t)}}{\cosh^n(t)} dt\nonumber\\&=& \int_0^{\pi/2} \cos^{n-1}(\theta) e^{-z\sec(\theta)}d\theta\,,
\end{eqnarray}
where $K_n(z)$ are modified Bessel functions and the second   equality is a well-known integral representation~\cite{abramowitz+stegun}. The last equality is obtained by the transformation $\cos(\theta) = \text{sech}(t)$. Using this form of the integrals $\mathcal{I}_n$, one can express the corrections to the fields in terms of these special functions. Below, we quote the results,
\begin{widetext}
    
\begin{eqnarray}
    \delta\vb{E}^{(1)}_{md}(\vb{r}) &=& \frac{\alpha}{6\pi^2 r^3} \Big\{(3\hat{\vb{r}}(\vb{m}\cdot \hat{\vb{r}}) - \vb{m})\Big(\text{Ki}_{0}(2m_f r) + 2m_f r \text{Ki}_{-1}(2m_f r)\nonumber \\
    &&- \frac{1}{2}\left[\text{Ki}_{2}(2m_f r) + \text{Ki}_{4}(2m_f r) + 2m_f r \text{Ki}_{1}(2m_f r) + 2m_f r\text{Ki}_{3}(2m_f r)\right]\Big)\nonumber\\
    &&+ 4m_f^2\vb{r}(\vb{m}\cdot \vb{r})\Big( \text{Ki}_{-2}(2m_f r) - \frac{1}{2}[\text{Ki}_{0}(2m_f r) + \text{Ki}_{2}(2m_f r)] \Big)\Big\}\,,
    \\
    \delta \vb{B}^{(1)}_{md}(\vb{r}) &=& \frac{\alpha}{6\pi^2 r^3} \Big\{(3\hat{\vb{r}}(\vb{m}\cdot \hat{\vb{r}}) - \vb{m})\Big(\text{Ki}_{0}(2m_f r) + 2m_f r \text{Ki}_{-1}(2m_f r)\nonumber \\
    &&- \frac{1}{2}\left[\text{Ki}_{2}(2m_f r) + \text{Ki}_{4}(2m_f r) + 2m_f r \text{Ki}_{1}(2m_f r) + 2m_f r\text{Ki}_{3}(2m_f r)\right]\Big)\nonumber\\
    &&+ 4m_f^2 r^2(\hat{\vb{r}}\times(\vb{\hat{r}}\times\vb{m}))\Big( \text{Ki}_{-2}(2m_f r) - \frac{1}{2}[\text{Ki}_{0}(2m_f r) + \text{Ki}_{2}(2m_f r)] \Big)\Big\}.
\end{eqnarray}

\end{widetext}

\bibliographystyle{JHEP}
\bibliography{cites}

\end{document}